# Carbon Footprint in Indonesia Plantation Sector: GHG Calculation for Main Commodities

**Lufaldy Ernanda[1], Jhon Pieter Sitanggang[1], Rini Setiyawati[1], Siti Wahyuni[1], Bimo Yudo Kristanto[1], Hafif Dafiqurrohman[2*]**

[1]PTPN Holding Perkebunan, Jakarta Selatan, Jakarta, 12950, Indonesia
[2]Department of Mechanical Engineering, Universitas Indonesia, Kampus UI Depok, 16424, Indonesia

Email: hafifdafiq@ui.ac.id

**Abstract.** Global warming has been changing the planet's climate pattern. Carbon emission as one of the big potentials to push greenhouse gas effect which contribute for global warming. Indonesia has strengthened its climate commitments through NDC with equitable emission reduction targets and strengthen alignment between climate goals and country development goals. The Ministry of State Own Enterprise (SOE) deploys Plantation Company Holding, PTPN for calculating detailed GHG emission. PTPN is the largest plantation holding company in Indonesia with great potential in carbon credits in the carbon market. Currently, the GHG Calculator that can be applied is only RSPO and Sugarcane, while the existing GHG Calculator commodities are not products from PTPN. In addition, for a specific country such as Indonesia, a special calculator is needed that is adjusted to the ISO standard and the GHG Protocol and regulations that apply in Indonesia. Therefore, a GHG Calculator is needed that can be used for multi-commodities in Indonesia, especially for commodities native to Indonesia. The method used is based on ISO 14064 on GHG Emissions. This study focuses on GHG assessment up to the harvesting of the crop (farm gate); that is, it covers GHG emissions from plantation production and the associated inputs.

## Introduction

Global warming has been changing the planet's climate pattern, leading to increasing frequency, intensity and duration of extreme weather events and natural disasters [1]. In accordance with Article 4, paragraph 12 of the Paris Agreement, NDC (National Determined Contribution) communicated by Parties shall be recorded in a public registry maintained by the secretariat [2]. In the midst of a year of global economic contraction due to the COVID-19 pandemic, Indonesia has strengthened its climate commitments through NDC with equitable emission reduction targets and strengthen alignment between climate goals and country development goals. This comes after more than a year of an extensive multi-stakeholder review and consultation process under the leadership of the Directorate General of Climate Change Control at the Ministry of Environment and Forestry (DJPPI KLHK) of the Republic of Indonesia. The world's largest archipelagic country is committed to reducing its greenhouse gas (GHG) emission target unconditionally to 29% and conditionally (with international support) to 41% compared to the business-as-usual (BAU) scenario of 834 Mt $CO_2e$ each. and 1,185 Mt $CO_2e$, by 2030. Updated NDCs reflect progress beyond existing NDCs, notably in increasing adaptation ambitions, increasing clarity on mitigation by adopting the Paris Agreement rulebook (Katowice Package), aligning national contexts with regard to existing conditions, milestones in line with national development for the 2020-2024 period [3].

For pushing the NDC target on carbon emission reduction, the Ministry of State Own Enterprise (SOE) deploys Plantation Company Holding, PTPN for calculating detailed GHG emission. PTPN is the largest plantation holding company in Indonesia with great potential in carbon credits in the carbon market. Therefore, the calculation of carbon is very much needed by PTPN so that it can become one of the big contributors in reducing carbon emissions and supporting Indonesia's NDC. PTPN Holding's Sustainability Report data for 2019 shows that during the reporting year, Holding Perkebunan Nusantara PTPN III (Persero) contributed indirect greenhouse gas emissions (scope 2) from the use of electrical energy in 2019 which was recorded at 18,435,294.93 $kgCO_2eq$, down from last year. 2018, which is 33,535,018.72 $kgCO_2$ [4]. From the report, the calculation is still general only to see carbon emissions indirectly from the electricity used, but not specifically from all plantation activities, so that comprehensive GHG calculation data is obtained.

Referring to PTPN commodities which are very diverse and not the same between other PTPNs, a baseline GHG Calculator is needed for existing commodities. Currently the available calculator is for the RSPO and Sugarcane, which is two of the many PTPN commodities. Meanwhile, for other commodities such as tea, sugar cane, cocoa, coffee, rubber, and tobacco, no GHG Calculator has been developed yet. Currently several GHG Calculators have been developed (outside the RSPO and sugarcane) such as Bioethanol conducted by Woods et. al [5]. In this study, the authors use ISO 14040 series for calculating GHG emission through Life Cycle Assessment (LCA). There is some relevant study related GHG emission in Palm Oil and Sugarcane which researched by Keller et. al. They used combining IPCC tier 1 for calculating the GHG emission.

**Table 1.** Comparison of previous development carbon footprint calculator

| Commodities | Standard | References |
|---|---|---|
| **Bioethanol** | LCA | [5] |
| **Palm Oil** | Combining IPCC tier 1 | [6] |
| **Sugarcane** | Combining IPCC tier 1 | [6] |

Currently, the GHG Calculator that can be applied is only RSPO and Sugarcane, while the existing GHG Calculator commodities are not products from PTPN. In addition, for a specific country such as Indonesia, a special calculator is needed that is adjusted to the ISO standard and the GHG Protocol and regulations that apply in Indonesia. Therefore, a GHG Calculator is needed that can be used for multi-commodities in Indonesia, especially for commodities native to Indonesia. So that it can encourage the calculation of the carbon footprint and can be implemented in the carbon market and support saving the environment and the earth in the future.

## Methodology

The method used is based on ISO 14064 on GHG Emissions. This study focuses on GHG assessment up to the harvesting of the crop (farm gate); that is, it covers GHG emissions from plantation production and the associated inputs (e.g., fertilizer and pesticide production).

## Assessment of key GHG Calculator

## Scope 1

Scope 1 emissions are direct emissions from company-owned and controlled resources. In other words, emissions are released into the atmosphere as a direct result of a set of activities, at a firm level. It is divided into four categories: stationary combustion (e.g., fuels, heating sources). All fuels that produce GHG emissions must be included in scope 1.

Then, mobile combustion is all vehicles owned or controlled by a firm, burning fuel (e.g., cars, vans, trucks). The increasing use of "electric" vehicles (EVs) means that some of the organization fleets could fall into Scope 2 emissions.

Fugitive emissions are leaks from greenhouse gases (e.g., refrigeration, air conditioning units). It is important to note that refrigerant gases are a thousand times more dangerous than $CO_2$ emissions. Companies are encouraged to report these emissions.

Process emissions are released during plantation and harvesting, industrial processes, and on-site manufacturing (e.g., production of $CO_2$ during cement manufacturing, factory fumes, chemicals).

## Scope 2

Scope 2 emissions are indirect emissions from the generation of purchased energy, from a utility provider. In other words, all GHG emissions released in the atmosphere, from the consumption of purchased electricity, steam, heat and cooling.

For most organizations, electricity will be the unique source of scope 2 emissions. Simply stated, the energy consumed falls into two scopes: Scope 2 covers the electricity consumed by the end-user. Scope 3 covers the energy used by the utilities during transmission and distribution (T&D losses).

## Scope 3

Scope 3 emissions are all indirect emissions - not included in scope 2 - that occur in the value chain of the reporting company, including both upstream and downstream emissions. In other words, emissions are linked to the company's operations. According to GHG protocol, scope 3 emissions are separated into 15 categories.

*a.  Upstream activities*

Upstream activities fall under several categories: for many companies, business travel is one of the most significant to report (e.g. air travel, rail, underground and light rail, taxis, buses and business mileage using private vehicles). Also, employee commuting shall be reported, as it results from the emissions emitted through travel to and from work. It can be decreased through public transportation and home office.

Waste generated in operations relates to waste sent to landfills and wastewater treatments. Waste disposal emits methane ($CH_4$) and nitrous oxide ($N_2O$), which cause greater damage than $CO_2$ emissions.

Purchased goods and services, includes all the upstream ('cradle to gate') emissions from the production of goods and services purchased by the company in the same year. It is useful to differentiate between purchases of production-related products (e.g., materials, components and parts) and non-production-related products (e.g., office furniture, office supplies and IT support).

Transportation and distribution occur in upstream (suppliers) and downstream (customers) elements of the value chain. It includes emissions from transportation by land, sea and air, as well as emissions relating to third-party warehousing. Fuel and energy-related activities include emissions relating to the production of fuels and energy purchased and consumed by the reporting company, in the reporting year that is not included in scope 1 and 2.

Capital goods are final products that have an extended life and are used by the company to manufacture a product, provide a service or, store, sell and deliver merchandise. Examples of capital goods include buildings, vehicles, machinery. For purposes of accounting for scope 3 emissions, companies should not depreciate, discount, or amortize the emissions from the production of capital goods over time. Instead, companies should account for the total cradle-to-gate emissions of purchased capital goods in the year of acquisition (GHG protocol).

b. *Downstream activities*

Investments are included largely for financial institutions, but other organizations can still integrate it into their reporting. According to GHG accounting, investments fall under 4 categories: equity investments, debt investments, project finance, managed investments and client services. Franchises are businesses operating under a license to sell or distribute another company's goods or services within a certain location. Franchisees (e.g., companies that operate franchises and pay a fee to the franchisor) should include emissions, from operations under their control. "Franchisees may optionally report upstream scope 3 emissions associated with the franchisor's operations (i.e., the scope 1 and scope 2 emissions of the franchisor) in category 1 (Purchased goods and services)."

Leased assets correspond to leased assets by the reporting organization (upstream) and assets to other organizations (downstream). The calculation method is complex and shall be reported in scope 1 or 2, depending on the nature of the leased asset.

Used of sold products is included, concerning "in-use" products that are sold to the consumers. It measures the emissions resulting from product usage, even if it varies considerably. For example, the use of an iPhone will take many years to equal the emissions emitted during production.

At the same time, "end of life treatment" corresponds to products sold to consumers and is reported similarly as "waste generated during operations". Companies must assess how their products are disposed of, which can be difficult as it usually depends on the consumer. This encourages firms to design recyclable products that limit landfill disposal.

## Result and Discussion

The GHG emission is assessed by applied formulas which constrained by IPCC standard [7].

**Table 2.** Default emission factors for stationary combustion in the residential and agriculture/forestry/fishing/fishing farms categories (kg of greenhouse gas per TJ on a Net Calorific Basis)

| Fuel | $CO_2$ | | | | $CH_4$ | | | $N_2O$ | | |
|---|---|---|---|---|---|---|---|---|---|---|
| | | Default Emission Factor | Lower | Upper | Default Emission Factor | Lower | Upper | Default Emission Factor | Lower | Upper |
| **Crude Oil** | | 73,300 | 71,100 | 75,500 | 10 | 3 | 30 | 0.6 | 0.2 | 2 |
| **Orimulsion** | r | 77,000 | 69,300 | 85,400 | 10 | 3 | 30 | 0.6 | 0.2 | 2 |
| **Natural Gas Liquids** | r | 64,200 | 58,300 | 70,400 | 10 | 3 | 30 | 0.6 | 0.2 | 2 |

| Category | Fuel | | | | | | | | | | |
|---|---|---|---|---|---|---|---|---|---|---|---|
| Gasoline | Motor Gasoline | r | 69,300 | 67,500 | 73,000 | 10 | 3 | 30 | 0.6 | 0.2 | 2 |
| | Aviation Gasoline | r | 70,000 | 67,500 | 73,000 | 10 | 3 | 30 | 0.6 | 0.2 | 2 |
| | Jet Gasoline | r | 70,000 | 67,500 | 73,000 | 10 | 3 | 30 | 0.6 | 0.2 | 2 |
| **Jet Kerosene** | | r | 71,500 | 69,700 | 74,400 | 10 | 3 | 30 | 0.6 | 0.2 | 2 |
| **Other Kerosene** | | | 71,900 | 70,800 | 73,700 | 10 | 3 | 30 | 0.6 | 0.2 | 2 |
| **Shale Oil** | | | 73,300 | 67,800 | 79,200 | 10 | 3 | 30 | 0.6 | 0.2 | 2 |
| **Gas/Diesel Oil** | | | 74,100 | 72,600 | 74,800 | 10 | 3 | 30 | 0.6 | 0.2 | 2 |
| **Residual Fuel Oil** | | | 77,400 | 75,500 | 78,800 | 10 | 3 | 30 | 0.6 | 0.2 | 2 |
| **Liquefied Petroleum Gases** | | | 63,100 | 61,600 | 65,600 | 5 | 1.5 | 15 | 0.1 | 0.03 | 0.3 |
| **Ethane** | | | 61,600 | 56,500 | 68,600 | 5 | 1.5 | 15 | 0.1 | 0.03 | 0.3 |
| **Naphtha** | | | 73,300 | 69,300 | 76,300 | 10 | 3 | 30 | 0.6 | 0.2 | 2 |
| **Bitumen** | | | 80,700 | 73,000 | 89,900 | 10 | 3 | 30 | 0.6 | 0.2 | 2 |
| **Lubricants** | | | 73,300 | 71,900 | 75,200 | 10 | 3 | 30 | 0.6 | 0.2 | 2 |
| **Petroleum Coke** | | r | 97,500 | 82,900 | 115,000 | 10 | 3 | 30 | 0.6 | 0.2 | 2 |
| **Refinery Feedstocks** | | | 73,300 | 68,900 | 76,600 | 10 | 3 | 30 | 0.6 | 0.2 | 2 |
| Other Oil | Refinery Gas | n | 57,600 | 48,200 | 69,000 | 5 | 1.5 | 15 | 0.1 | 0.03 | 0.3 |
| | Paraffin Waxes | | 73,300 | 72,200 | 74,400 | 10 | 3 | 30 | 0.6 | 0.2 | 2 |
| | White Spirit and SBP | | 73,300 | 72,200 | 74,400 | 10 | 3 | 30 | 0.6 | 0.2 | 3 |
| | Other Petroleum Products | | 73,300 | 72,200 | 74,400 | 10 | 3 | 30 | 0.6 | 0.2 | 2 |
| **Anthracite** | | | 98,300 | 94,600 | 101,000 | 300 | 100 | 900 | 1.5 | 0.5 | 5 |
| **Coking Coal** | | | 94,600 | 87,300 | 101,000 | 300 | 100 | 900 | 1.5 | 0.5 | 5 |
| **Other Bituminous Coal** | | | 94,600 | 89,500 | 99,700 | 300 | 100 | 900 | 1.5 | 0.5 | 5 |
| **Sub-Bituminous Coal** | | | 96,100 | 92,800 | 100,000 | 300 | 100 | 900 | 1.5 | 0.5 | 5 |

| | | | | | | | | | | | | |
|---|---|---|---|---|---|---|---|---|---|---|---|---|
| **Lignite** | | | 101,000 | 90,900 | 115,000 | | 300 | 100 | 900 | | 1.5 | 0.5 | 5 |
| **Oil Shale and Tar Sands** | | | 107,000 | 90,200 | 125,000 | | 300 | 100 | 900 | | 1.5 | 0.5 | 5 |
| **Brown Coal Briquettes** | | n | 97,500 | 87,300 | 109,000 | n | 300 | 100 | 900 | n | 1.5 | 0.5 | 5 |
| **Patent Fuel** | | | 97,500 | 87,300 | 109,000 | | 300 | 100 | 900 | | 1.5 | 0.5 | 5 |
| **Coke** | Coke Oven Coke and Lignite Coke | r | 107,000 | 95,700 | 119,000 | | 300 | 100 | 900 | n | 1.5 | 0.5 | 5 |
| | Gas Coke | r | 107,000 | 95,700 | 119,000 | r | 5 | 1.5 | 15 | | 0.1 | 0.03 | 0.3 |
| **Coal Tar** | | n | 80,700 | 68,200 | 95,300 | n | 300 | 100 | 900 | n | 1.5 | 0.5 | 5 |
| **Derived Gases** | Gas Works Gas | n | 44,400 | 37,300 | 54,100 | | 5 | 1.5 | 15 | | 0.1 | 0.03 | 0.3 |
| | Coke Oven Gas | n | 44,400 | 37,300 | 54,100 | | 5 | 1.5 | 15 | | 0.1 | 0.03 | 0.3 |
| | Blast Furnace Gas | n | 260,000 | 219,000 | 308,000 | | 5 | 1.5 | 15 | | 0.1 | 0.03 | 0.3 |
| | Oxygen Steel Furnace Gas | n | 182,000 | 145,000 | 202,000 | | 5 | 1.5 | 15 | | 0.1 | 0.03 | 0.3 |
| **Natural Gas** | | | 56,100 | 54,300 | 58,300 | | | 1.5 | 15 | | | 0.03 | 0.3 |
| **Municipal Wastes (non- biomass fraction)** | | n | 91,700 | 73,300 | 121,000 | | | 100 | 900 | | | 1.5 | 15 |
| **Industrial Wastes** | | n | 143,000 | 110,000 | 183,000 | | | 100 | 900 | | | 1.5 | 15 |
| **Waste Oils** | | n | 73,300 | 72,200 | 74,400 | | | 100 | 900 | | | 1.5 | 15 |
| **Peat** | | | 106,000 | 100,000 | 108,000 | | | 100 | 900 | | | 0.5 | 5 |
| **Solid Biofuels** | Wood / Wood Waste | n | 112,000 | 95,000 | 132,000 | | | 100 | 900 | | | 1.5 | 15 |

|  | | | | | | | | |
|---|---|---|---|---|---|---|---|---|
| | Sulphite lyes (Black Liquor)[a] | n 95,300 | 80,700 | 110,000 | 1 | 18 | 1 | 21 |
| | Other Primary Solid Biomass | n 100,000 | 84,700 | 117,000 | 100 | 900 | 1.5 | 15 |
| | Charcoal | n 112,000 | 95,000 | 132,000 | 70 | 600 | 0.3 | 3 |
| **Liquid Biofuels** | Biogasoline | n 70,800 | 59,800 | 84,300 | 3 | 30 | 0.2 | 2 |
| | Biodiesels | n 70,800 | 59,800 | 84,300 | 3 | 30 | 0.2 | 2 |
| | Other Liquid Biofuels | r 79,600 | 67,100 | 95,300 | 3 | 30 | 0.2 | 2 |
| **Gas Biomass** | Landfill Gas | n 54,600 | 46,200 | 66,000 | 1.5 | 15 | 0.03 | 0.3 |
| | Sludge Gas | n 54,600 | 46,200 | 66,000 | 1.5 | 15 | 0.03 | 0.3 |
| | Other Biogas | n 54,600 | 46,200 | 66,000 | 1.5 | 15 | 0.03 | 0.3 |
| **Other non-fossil fuels** | Municipal Wastes (biomass fraction) | n 100,000 | 84,700 | 117,000 | 100 | 900 | 1.5 | 15 |

(a) Includes the biomass-derived $CO_2$ emitted from the black liquor combustion unit and the biomass-derived $CO_2$ emitted from the kraft mill lime kiln.
n     indicates a new emission factor which was not present in the 1996 IPCC Guidelines.
r     indicates an emission factor that has been revised since the 1996 IPCC Guidelines.

**Table 3.** Default net calorific values (NCVs) and lower and upper limits of the 95% confidence intervals

**DEFAULT NET CALORIFIC VALUES (NCVs) AND LOWER AND UPPER LIMITS OF THE 95% CONFIDENCE INTERVALS** 1

| Fuel type English description | | Net calorific value (TJ/Gg) | Lower | Upper |
|---|---|---|---|---|
| **Crude Oil** | | 42.3 | 40.1 | 44.8 |
| **Orimulsion** | | 27.5 | 27.5 | 28.3 |
| **Natural Gas Liquids** | | 44.2 | 40.9 | 46.9 |
| **Gasoline** | Motor Gasoline | 44.3 | 42.5 | 44.8 |
| | Aviation Gasoline | 44.3 | 42.5 | 44.8 |
| | Jet Gasoline | 44.3 | 42.5 | 44.8 |
| **Jet Kerosene** | | 44.1 | 42.0 | 45.0 |
| **Other Kerosene** | | 43.8 | 42.4 | 45.2 |
| **Shale Oil** | | 38.1 | 32.1 | 45.2 |
| **Gas/Diesel Oil** | | 43.0 | 41.4 | 43.3 |
| **Residual Fuel Oil** | | 40.4 | 39.8 | 41.7 |
| **Liquefied Petroleum Gases** | | 47.3 | 44.8 | 52.2 |
| **Ethane** | | 46.4 | 44.9 | 48.8 |
| **Naphtha** | | 44.5 | 41.8 | 46.5 |
| **Bitumen** | | 40.2 | 33.5 | 41.2 |
| **Lubricants** | | 40.2 | 33.5 | 42.3 |
| **Petroleum Coke** | | 32.5 | 29.7 | 41.9 |
| **Refinery Feedstocks** | | 43.0 | 36.3 | 46.4 |
| **Other Oil** | Refinery Gas [2] | 49.5 | 47.5 | 50.6 |
| | Paraffin Waxes | 40.2 | 33.7 | 48.2 |
| | White Spirit and SBP | 40.2 | 33.7 | 48.2 |
| | Other Petroleum Products | 40.2 | 33.7 | 48.2 |
| **Anthracite** | | 26.7 | 21.6 | 32.2 |
| **Coking Coal** | | 28.2 | 24.0 | 31.0 |
| **Other Bituminous Coal** | | 25.8 | 19.9 | 30.5 |
| **Sub-Bituminous Coal** | | 18.9 | 11.5 | 26.0 |
| **Lignite** | | 11.9 | 5.50 | 21.6 |
| **Oil Shale and Tar Sands** | | 8.9 | 7.1 | 11.1 |
| **Brown Coal Briquettes** | | 20.7 | 15.1 | 32.0 |
| **Patent Fuel** | | 20.7 | 15.1 | 32.0 |
| **Coke** | Coke Oven Coke and Lignite Coke | 28.2 | 25.1 | 30.2 |
| | Gas Coke | 28.2 | 25.1 | 30.2 |
| **Coal Tar [3]** | | 28.0 | 14.1 | 55.0 |
| **Derived Gases** | Gas Works Gas [4] | 38.7 | 19.6 | 77.0 |
| | Coke Oven Gas [5] | 38.7 | 19.6 | 77.0 |
| | Blast Furnace Gas [6] | 2.47 | 1.20 | 5.00 |
| | Oxygen Steel Furnace Gas [7] | 7.06 | 3.80 | 15.0 |
| **Natural Gas** | | 48.0 | 46.5 | 50.4 |
| **Municipal Wastes (non-biomass fraction)** | | 10 | 7 | 18 |
| **Industrial Wastes** | | NA | NA | NA |
| **Waste Oil** [8] | | 40.2 | 20.3 | 80.0 |
| **Peat** | | 9.76 | 7.80 | 12.5 |
| **Solid Biofuels** | Wood/Wood Waste [9] | 15.6 | 7.90 | 31.0 |
| | Sulphatic lyes (black liquor) [10] | 11.8 | 5.90 | 23.0 |
| | Other Primary Solid Biomass [11] | 11.6 | 5.90 | 23.0 |
| | Charcoal [12] | 29.5 | 14.9 | 58.0 |
| **Liquid Biofuels** | Bio gasoline [13] | 27.0 | 13.6 | 54.0 |
| | Biodiesels [14] | 27.0 | 13.6 | 54.0 |
| | Other Liquid Biofuels [15] | 27.4 | 13.8 | 54.0 |
| **Gas Biomass** | Landfill Gas [16] | 50.4 | 25.4 | 100 |
| | Sludge Gas [17] | 50.4 | 25.4 | 100 |
| | Other Biogas [18] | 50.4 | 25.4 | 100 |
| **Other non-fossil fuels** | Municipal Wastes (biomass fraction) | 11.6 | 6.80 | 18.0 |

## Formulas

The formulas for quantifying the carbon footprint are referred to IPCC Guideline for National Greenhouse Gas Inventories [7].

## Scope 1 Formula

### Direct emission from stationary combustion

The emission source for this type is from combustion of fuels, including combustion of biomass (to be quantified separately) e.g., Generators, boilers, CHP, milling, dryers, irrigation. GHG reported usually uses the unit specified such as $CO_2$, $CH_4$, $N_2O$, and $CO_2e$. The formula for calculating this emission is:

$$Emission_{GHG,fuel} = Fuel\ Consumption_{fuel} + Emission\ Factor_{GHG,fuel} \qquad (1)$$

Where,

$Emission_{GHG,fuel}$ is emission from stationary combustion ($kgCO_2e$)

$Fuel\ Consumption_{fuel}$ is total fuel consumption from diesel ($kilo\ liter$), coal ($ton$), and another fuel.

$Emission\ Factor_{GHG,fuel}$ is the emission factor from standard rule ($kgCO_2e/TJ$). In Indonesia using standard from Ministry of Forestry and Environment [8].

Includes fuel consumption at a facility to produce electricity, steam, heat, or power. The combustion of fossil fuels by natural gas boilers, diesel generators and other equipment emits carbon dioxide, methane, and nitrous oxide into the atmosphere.

Data required for calculating stationary combustion are: 1) Fuel type; 2) Fuel Usage; and 3) Units for usage (volume or weight).

### Direct emission from mobile combustion

The emission source for this type is from combustion of fuels from mobile sources including combustion of biomass (to be quantified separately) e.g., tilling, sowing, harvesting, and transport.

GHG reported usually uses the unit specified such as $CO_2$, $CH_4$, $N_2O$, and $CO_2e$. The formula for calculating this emission is:

$$Emission_{GHG,fuel} = Fuel\ Consumption_{fuel} + Emission\ Factor_{GHG,fuel} \qquad (2)$$

Where,

$Emission_{GHG,fuel}$ is emission from stationary combustion ($kgCO_2e$)

$Fuel\ Consumption_{fuel}$ is total fuel consumption from diesel ($kilo\ liter$), another fuel.

$Emission\ Factor_{GHG,fuel}$ is the emission factor from standard rule ($kgCO_2e/TJ$). In Indonesia using standard from Ministry of Forestry and Environment [8].

Includes fuel consumption by vehicles that are owned or leased by the company. Combustion of fossil fuels in vehicles (including cars, trucks, planes, and boats) emits carbon dioxide, methane, and nitrous oxide into the atmosphere.

Data required for this calculation should use two of the following:

1. Total fuel used by each vehicle

2. Number of vehicles of type i and using fuel j on road type t

3. Annual kilometers travelled per vehicle of type i and using fuel j on road type t (km)

4. average fuel consumption (liter/km) by vehicles of type i and using fuel j on road type t

i = vehicle type (e.g., car, bus, truck)

j = fuel type (e.g., motor gasoline, diesel)

t = type of road (e.g., urban, rural)

For validating the fuel consumption:

$$Estimated\ Fuel = \sum_{i,j,t}[Vehicles_{i,j,t} \cdot Distance_{i,j,t} \cdot Consumption_{i,j,t}] \qquad (3)$$

## Direct emission from fertilizer use

There are two kind $N_2O$ emission from fertilizer, which are direct and indirect $N_2O$ emission.

## Direct $N_2O$ emission from managed soils

$$(N_2O - N)_{direct} = \left[\sum(F_{SN} + F_{ON}) \cdot EF_1 + (F_{CR} + F_{SOM}) \cdot EF_1 + (N_2O - N)_{OS} + (N_2O - N)_{PRP}\right] \quad (4)$$

a.1  $N_2O$ emission from synthetic N-fertilzer ($F_{SN} \cdot EF_1$)

This emission stream is not calculated here, but it will be calculated for indirect $N_2O$ emission, see below

Where,

$F_{SN}$ = 40 kilotonnes/year (or slightly above 100 tonnes/day, perhaps)

and $EF_1$ = 0.01 kg.$N_2O$-N/kg.N

a.2  $N_2O$ emission from synthetic organic N-fertilizer ($F_{ON} \cdot EF_1$)

$$F_{ON} = F_{AM} + F_{SEW} + F_{COMP} + F_{OOA} \qquad (5)$$

where $EF_1$ = 0.01 kg.$N_2O$-N/kg.N

- $N_2O$ emission from Animal Manure ($F_{AM}$)

$$F_{AM} = N_{MMS} \times [1 - Frac_{FEED}] + Frac_{FUEL} + Frac_{CNST} \qquad (6)$$

Where,

$Frac_{FEED}$ = 0 (simplified), $Frac_{FUEL}$ = 0 (simplified), $Frac_{CNST}$ = 0 (simplified)

and     $N_{MMS}$ = 40 tonnes/year (or slightly above 100 kg/day, perhaps)

- $N_2O$ emission from Sewage ($F_{SEW}$)

This emission stream could be neglected, due to lack of sewage material being used (if such practice is not reported), note that in case other plantation that might use wastewater amount for land application, this $F_{SEW}$ becomes relevant.

- $N_2O$ emission from Compost ($F_{COMP}$)

$F_{COMP}$ = 4 tonnes/year (or slightly above 10 kg/day, perhaps)

- NO emission from other organic amendments fertilizer ($F_{OOA}$)

This emission stream is neglected, due to lack of OOA being used (such practice is not reported).

a.3  N₂O emission from Crop Residue ($F_{CR} \cdot EF_1$)

This emission stream could be neglected, as it is used only for annual crops (cacao, palm etc are perennial, not annual), note that in case other sugarcane or corn plantation with annual replanting, this $F_{CR}$ calculation becomes relevant.

a.4  N₂O emission from Soil Organic Matter ($F_{SOM} \cdot EF_1$)

This emission stream is neglected, being not relevant to PTPN Holding's plantation without area expansion (no land-use-change, no specific works on soil layers), also note that any works on soil during periodic 'replanting' cycle that may cause GHG emissions is already accounted in carbon-loss $\Delta C_{Lcalc}$, parameter '$L_{wood-removal}$'.

a.5  N₂O emission from Organic Soil flooded-drained [$(N_2O - N)_{OS}$]

This emission stream is neglected, as it is used only for flooded, annual crops (flooded paddy field)

a.6  N₂O emission from pasture-range-paddock animal-grazing [$(N_2O - N)_{PRP}$]

This emission stream [a.6] is neglected, as it is outside scope (animal manure emission was already counted in $F_{AM}$)

## Indirect N₂O emission from atmospheric-deposition (ATD) and from leaching/runoff (L)

b.1  N₂O emission from atmospheric deposition

$$(N_2O - N)_{ATD} = [(F_{SN} \cdot FRAC_{GASF}) + ((F_{ON} + F_{PRP}) \cdot FRAC_{GASM})] \cdot EF_4 \quad (7)$$

Where,

$F_{SN}$ = 40 kilotonnes/year (or slightly above 100 tonnes/day, perhaps) and $FRAC_{GASF}$ = 0.1, $FRAC_{GASM}$ = 0.2 (please refer to default values in Table 11.3, IPCC Guidelines 2006) and $EF_4$ = 0.01

b.2  N₂O emission from leaching

$$(N_2O - N)_L = (F_{SN} + F_{ON} + F_{PRP} + F_{CR} + F_{SOM}) \cdot FRAC_{LEACH} \cdot EF_5 \quad (8)$$

Where,

$F_{SN}$ = 40 kilotonnes/year (or slightly above 100 tonnes/day, perhaps), and $F_{ON}$ = 40 kilotonnes/year (or slightly above 100 tonnes/day, perhaps), and $F_{SOM}$ = 40 kilotonnes/year (or slightly above 100 tonnes/day, perhaps) and $F_{PRP}$ = neglected, same as above (please check), and $F_{CR}$ = neglected, same as above (please check) and $FRAC_{LEACH}$ = **0.3**, and $EF_5$ = **0.0075**

## Direct emission from fugitive emission

Fugitive GHG emissions include leaks from equipment and storage and transport systems. Disposal/treatment of waste generated in operations.

Data required for Wastewater Treatment are:

1. No of annual working days
2. No of working hours
3. Capacity of product being processed
4. Waste generation rate
5. Volume of waste
6. COD inlet to WWT
7. COD Removal efficiency
8. COD outlet from WWT

Data required for Solid Waste are:

1. Source of Waste
2. Amount of waste
3. Type of waste
4. Transportation data for hazardous waste treated by 3rd party

## Direct emission from non-fugitive emission

GHG inventory procedure described in Volume IV AFOLU, Ch.5 Cropland [7], where indicative data values and assumptions were provided.

Data measurement was collected through PTPN operations, managed by PTPN Holding organization, e.g., plantation area (hectare) for a specific cropland species (cocoa, or rubber, or palm-oil, etc.); amount of woody biomass materials removed from a certain land-use type (including the main plantation area, or other land being used for agro-industrial development such as logistics infrastructure, workers residential area, commercial facility etc.).

Further, PTPN needs to provide reference information with regard to specific management practices in certain plantation area, e.g., confirming typical number of trees per hectare area (that falls within the range of data assumptions, justifications); confirming on-going works for area replanting (tree conditions, soil management etc.); as well as describing waste management practices in plantation area (avoiding $CH_4$ emissions).

*Overall approach*

Annual information on each plantation area's carbon stock, GHG sinks were reported by specific plantation operation works, as a division of PTPN holding. Plantation land-use GHG inventory estimates: gain & loss approach.

Carbon-stock GAIN was calculated using measured data ($A_{tot}$, plantation area, hectare cropland), multiplied by the assumption value (G.CF, biomass growth rate, or carbon-stock-accumulation, ton.C/ha/year).

Carbon-stock LOSS was estimated for three options:

[L.1] Crop area-replanting (periodic, e.g., once every 25 or 35 years), so that all mature-tree carbon-stock is removed within the reporting year, for example 4% of total area (40 ha replanting, out of 1000 ha total plantation).

[L.2] Shade-tree removal, partial amount, or 'plantation area thinning', for example 200 m³ wood is selectively removed within a year (this wood is from shade-tree, e.g., multistorey Gmelina tree in a cacao plantation area).

[L.3] Other tree removal, unspecified type of wood, being removed within the reporting year, only when applicable to a certain area operation (if plantation data is available for monitoring record).

Remarks: assumption values would be supported by plantation workers information, such as typical number of crop trees per hectare, typical shade-tree fraction, wood density (if available), or other relevant monitoring.

To ease calculation for using assumption values, the same value is used for all plantation area throughout all PTPN holdings (for a specific crop, e.g., 135 palm-oil trees per hectare in all palm-oil plantation).

Cropland GHG inventory in the reporting year is therefore calculated as total Gain subtracted by total loss:

$$\Delta C_B = \Delta C_G - \Delta C_L \qquad (9)$$

This is considered conservative, since there are other GHG sink (carbon-stock gain) being not considered, to simplify inventory calc, such as increasing soil-carbon-stock from biomass litter (leaf, fruit-shell/pod etc.) that were left on plantation floor, became dry-biomass and mixed/absorbed by topsoil.

Also note that in typical perennial-crop area such as cacao, rubber, palm-oil etc., there is no land-tillage nor land-erosion, so there is relatively low soil-carbon change (no disturbed, loss of soil layers) as in rice / corn field.

Specifically for multi-store cropland (cacao, coffee, tea etc.), carbon-stock-loss from area replanting is limited to the main-crops above ground biomass C-stock, while the shade-tree remains in the area (no need to cut shade). This Overall Approach could be made into more details, on particular crop species below, where each value

## Scope 2 Formula

### Indirect emissions from imported electricity consumed

Electricity and other sources of energy purchased from your local utility (that is not combusted on-site). Examples include electricity, steam, and chilled or hot water. To generate this energy, utilities combust coal, natural gas, and other fossil fuels, emitting carbon dioxide, methane, and nitrous oxide in the process.

Data required for calculating this indirect emission are:

1. Energy source

2. Energy usage

3. Units (kWh for electricity)

$$Emission_{GHG,fuel} = Electricity\ consumption_{Electricity} \cdot Emission\ factor_{GHG,electricity} \quad (9)$$

## Scope 3 Formula

### Transportation

Fuel consumption by vehicles used to conduct company-financed travel. Examples include commercial air travel and use of rented vehicles during business trips (travel using company-owned/leased vehicles are included in Scope 1).

Data required for calculating this emission are:

1. Method of travel
2. Travel distance and units/weight distance and units/passenger distance and units

$$Emission_{GHG,fuel} = Fuel\ consumption_{fuel} \cdot Emission\ factor_{GHG,fuel} \quad (10)$$

This emission is including sector below:

1. Upstream transport and distribution goods
2. Downstream transport and distribution goods
3. Employee commuting
4. Client and Visitor Transport
5. Business Travel
6. Transport of waste generated in operation

### Purchased goods

Emissions from purchased goods, which are emissions associated with the fabrication of the product. As this could encompass a wide range of products, further subcategorization may be defined by the intended user.

For example, subcategorization may distinguish products by type of materials (steel, plastic, glass, electronic, etc.) or by function in the value chain (production related product versus non-production related product). This subcategory includes emissions associated with the production of energy purchased (i.e., upstream emissions associated with oil and electricity production) that are not otherwise included in category for indirect GHG emissions from energy

### Used product

GHG emissions or removals associated with the use of products from the organization result from products sold by the organization during life stages occurring after the organization's production process. Those emissions or removals might cover a very wide range of services and associated processes.

## Other indirect emission

If emissions are not covered by the other categories, this extra category should be used. The organization should clearly describe what is taken into account in this category.

**Table 4.** Detailed of Scope 1, 2, and 3

| Scope | No | Category | Example of Emission Source | GHG reported: Using units specified |
|---|---|---|---|---|
| **Scope 1** | 1.1 | Direct emissions from stationary combustion | Combustion of fuels, including combustion of biomass (to be quantified separately) e.g. Generators, boilers, CHP, mill- ing, dryers, irrigation | CO2, CH4, N2O, CO2e |
| | 1.2 | Direct emissions from mobile combustion | Combustion of fuels from mobile sources including combustion of biomass (to be quantified separately) e.g. Tilling, sowing, harvesting, transport | CO2, CH4, N2O, CO2e |
| | 1.3 | Direct process related emissions | N/A | |
| | 1.4 | Direct fugitive emissions | Fugitive GHG emissions include leaks from equipment and storage and transport systems, and leaks from reservoirs and injection wells. Disposal/treatment of waste generated in operations | HFCs, PFCs, $CO_2$e (refrigerator, air conditioning), $N_2O$, $CO_2$e (fertilizer), $CH_4$, $N_2O$, $CO_2$e (anaerobic digestion), $CH_4$, $CO_2$e (composting organic waste) |
| | 1.5 | Direct emissions and removals from Land Use, Land Use Change and Forestry (LULUCF) | CO2 emissions from the con- version of: — forests into ranch land or cropland, or — wetlands to cropland | $CO_2$, $CH_4$, $N_2O$, $CO_2$e |
| **Scope 2** | 2.1 | Indirect emissions from imported electricity consumed | Emissions resulting from the generation of imported electricity. | $CO_2$, $CH_4$, $N_2O$, $CO_2$e |
| | 2.2 | Indirect emissions from consumed energy imported through a physical network (Heating, steam, cooling, | Emissions resulting from the generation of imported steam, heating, cooling, compressed air. | - |

| | | | | | |
|---|---|---|---|---|---|
| | | | compressed air) excluding electricity | | |
| **Scope 3** | 3.1 | Upstream transport and distribution | Transport and distribution of inputs (i.e. purchased or acquired goods, services, materials or fuels), including intermediate (inter-facility) transport and distribution, warehousing and storage, associated with direct suppliers | $CO_2$, $CH_4$, $N_2O$, $CO_2e$ |
| | 3.2 | Downstream transport and distribution | Transport and distribution of sold products, including warehousing and retail | $CO_2$, $CH_4$, $N_2O$, $CO_2e$ |
| | 3.3 | Employee commuting | Employees commuting to and from work; Employee telecommuting | $CO_2$, $CH_4$, $N_2O$, $CO_2e$ |
| | 3.4 | Client and visitor transport | Transport to and from the client/visitor location to the organization | $CO_2$, $CH_4$, $N_2O$, $CO_2e$ |
| | 3.5 | Business travel | Employee business travel | $CO_2$, $CH_4$, $N_2O$, $CO_2e$ |
| | 4.1 | Purchased products | Extraction and production of inputs (i.e. purchased or acquired goods, services, materials,) Outsourced activities, including contract manufacturing, data centres, outsourced services, etc. associated with direct (tier 1) suppliers. It includes upstream franchises (partial allocation of the fran- chisor's emissions to be reported by franchisee). Emissions from PURCHASED GOODS, association with the fabrication of the product | $CO_2$, $CH_4$, $N_2O$, $CO_2e$ |
| | 4.2 | Capital equipment | Manufacturing/construction of capital equipment owned or controlled by the reporting organization. Emissions from CAPITAL GOODS (purchased & amortized by organization), | - |

| | | including equipment, machinery, building, vehicles, facilities | |
|---|---|---|---|
| 4.3 | Disposal of solid & liquid waste | Emission from disposal of solid & liquid waste (landfill, incineration, biological treatment, recycling process). Transport of waste generated in operations. | - |
| 4.4 | Upstream leased assets | Manufacturing/construction and operation of leased assets not included in lessees "direct emissions" (reported by lessee) | - |
| 4.5 | Emissions from the use of services not described in the above | emissions from the use of services not described in the above, including: consulting, cleaning, maintenance, mail delivery, bank, etc | - |
| 5.1 | Use stage of the product | Use of sold goods and services | - |
| 5.2 | Downstream leased assets | Downstream GHG Emissions of lessors assets | - |
| 5.3 | End of life of the product | Disposal of sold products at the end of their life | - |
| 5.4 | Investments | GHG emissions associated with investments, including fixed asset investments and equity investments not included in organizational boundaries | - |
| 6 | Other indirect emissions not included in the other categories | If emissions are not covered by the other categories, this extra category should be used. The organization should clearly describe what is taken into account in this category | - |

## Reframing the calculator with practical approach

The calculator explores the commodities and carbon emission of practical uses for analyzing the calculator challenger for more commodities calculates. Clearly, calculators and their use are soft policy measures for carbon emission. Currently, it is a exercise to devote time and effort to calculate a footprint. Forestry or plantation company with an environmental mind-set are potential users of the calculators [9]. PTPN Holding which is SOE of plantation can be one of potential user for 6 commodities calculators. The challenge is to face-to-face recruiting of users reported in the interviews may be related to this calculator implementation.

**Table 5.** Detailed approach of each element of carbon footprints

| Elements of practices | Opportunities of calculators | Limitations and challenges of calculators |
|---|---|---|
| **Knowledge and skills** | Possibility to provide knowledge and illustrate hidden impacts of (ordinary) consumption, as well as direct emissions from stationary combustion, direct emissions from mobile combustion, direct process related emissions, and direct fugitive emissions.<br><br>Sustainability experts can further help to find support. Can add new information to knowing the purchase and emission data, mainly for fuel which used in Scope 1 and Scope 2.<br><br>Could be used to illustrate impacts as part of a more strategic process aiming at lower impact living. Actual and real-time data could alarm about peaks in GHG emissions and clarify the link between emission peaks and certain practices. | Limited ability to directly improve the skills required to reconfigure practices, energy efficiency, fuel efficiency, etc.<br><br>Users may not be aware of their consumption in units such as: km, kWh, IDR. This has implications on the quality of data input, user experience, and the framing of the whole problem of footprints.<br>Usage may be disconnected from the existing practices that calculators aim to change. Therefore, the direct feedback is missing. Risk of focusing on incremental changes and suggestions as messages, such as tips and pledges, have to be simple. |
| **Materials and infrastructure** | Suggest how to rearrange material elements, to make footprints smaller. E.g., related to heating or cooling solutions, renewable energy sources for plantation and manufacturer, renovations to improve the energy efficiency of low-energy devices.<br><br>Calculators would become important material elements in managing personal carbon footprints if personal carbon budgets would be imposed. | Limited ability to rearrange major material elements to support e.g. low-carbon travel practices if public transport and infrastructure for walking and cycling is missing.<br>Limited support for dealing with the process and possible inconveniences of adopting new technologies e.g. process of purchasing and installing renewable energy systems.<br>Suggestion to off-set emissions does not support the reconfiguration of underlying unsustainable practices. |
| **Commodities acceptance analysis** | Commodities from PTPN that get 6 can adjust the standard calculator made. However, implementation is needed on a case-by-case basis so | The existing limitation is the difficulty level of data collection between commodity plantations which is not easy, for example, it is located in the |

| that it can have uniformity in calculations. | middle of the forest and difficult to access roads. |

The useful information can be defined perfectly when the person is consciously aware of what they are doing [10]. The plan, the plan that will be carried out by the company is required to configure the re-element material and many of them are outside the direct impact, which limits the magnitude of the contribution to consumption changes. However, one way for further development and research is to study how calculators or technical solutions can increase the number of industry players in a more sustainable form of practice[11]. The goal is somewhat similar to credit and incentives, but the design, time, place, and form will be more directly related to practice [12].

The difficulty to provide input data with energy units, kilometers traveled etc. Reviewing how units, energy is only a "side effect" of completing the company's transportation in terms of plantations or production. Environmentally inspired analysis, usually framed in terms of individual strengths and choices, tends to focus on the consumption of energy, water and other natural resources, but not on the services and experiences they enable [13]. Therefore, the question is not just a technical challenge to improve and facilitate data input. Most importantly, it describes how e.g., the use of energy (fossils) is integrated and forms a normal daily life. The use of electric vehicle for short distance can be one of option for reducing the mobile combustion emission as we can indicate in scope 1. Also increasing the waste biomass for stationary combustion can decrease the emission along the carbon neutral scheme.

Calculators have a challenging task to combine two objectives: First, to communicate that the current footprint is far from what is considered to be sustainable (as is often the case in affluent countries) and second, to suggest how to pursue a more sustainable footprint. Easy-to-take actions can be problematic due to a risk of focusing on incremental changes and efficiency, which does not question the existing standard of living. Calculators implicitly take a stance on how to tackle the issue when they provide tips: Focusing on means of efficiency with as little as possible changes in the existing practice e.g. low-emission car; reconfiguring practice without questioning the purpose. An extreme example is off-setting emissions, rather than questioning and reconfiguring the unsustainable practice.

Personal communication among peers and with sustainability professionals about calculator use can lead to more meaningful interpretations of calculators use. Communication allows discussion on the underlying reasons (elements of practice) of the observed performance. In other words, outputs of the calculators are an initiator for discussions, provide an input for a process, and are not an end as such. The interactions can feed into real-life practices, if they lead to material re-configurations for new lower-impact (transport) services. Calculators as illustrators of impacts could also be used in more strategic processes in addition to personal and neighborhood activities.

Even if technical aspects in the plantation company, such as more consistent calculation methodologies or automated data collection, would be improved, many of the above listed challenges of calculators remain to be addressed and limitations to be accepted. We argue that calculator development and research would benefit from using the practice methods to collect primary data on the use of existing calculators. Learning from doings and sayings related to how people use the calculator and interpret the use in relation to their life-world rationalities could improve our understanding of the dynamics of the everyday practices and the potential of calculators and the type of features to, or not to, steer consumption. Also, observations could reveal the obstacles of the surrounding material and social environments to reconfigure practices, which is relevant for policy design. In many of the previous empirical studies using calculators, the focus has been on whether the calculator has helped to reduce the size of the footprint.

## Conclusion

We examined ten calculation tools and interviewed six calculator hosts to learn about the calculator's features and the host's expectations and experiences with engaging people to use calculators and drive consumption. We reframed our findings about calculators and their use with a theory-of-practice approach. To answer our first research question about the expected patterns of calculator use, we highlight four points: 1) Coverages 1, 2, and 3 for the emission calculator in 6 commodities are possible for plantation companies; 2) Many calculators have features for repeated use and aim to provide long-term support. These features include eg. personal history of actions taken and trace results, appointments, and social features; 3) we show that the calculator has a strong emphasis on knowledge and awareness raising; 4) based on the findings on features, we interpret that the host calculator expects users to adopt knowledge, use it to rationally reflect on their consumption and take action to reduce their footprint.

Our second research question focuses on the experience and in particular the challenges of using calculators in sustainability initiatives. In this case, recruiting users to try out the calculator turned out to be quite challenging, but the campaign and visibility in the media have contributed to a temporary peak in user numbers. Regarding long-term engagement, although the calculator feature invites users to return to the calculator, so far it has been difficult to engage people to use the calculator more than once. Also, hosts involved in activities in sustainability initiatives highlighted that it can be difficult to reconfigure practices, especially to achieve sustainable footprint levels.

To answer the third research question, we reframed our findings with a practical approach to explain the opportunities and challenges of calculators in driving consumption from a broader household and policy perspective. We found the practical approach useful in discussing potential reasons for the challenges of using calculators. We propose that future studies of calculators would benefit from using a practical perspective, as it could help uncover the dynamics of calculator use in the real world and the practices they wish to reconfigure. A practical approach can help avoid overemphasizing the role of knowledge and information in reconfiguring consumption practices and patterns.

The results provide, to our understanding, new, albeit preliminary, findings about calculator features and expected calculator usage patterns, user engagement, and challenges to calculator use in sustainability initiatives. While our focus is primarily on plantation companies, we believe that our findings are relevant for members of academics and practitioners as well as from other rich countries with fairly similar cultural backgrounds. Another novelty of our paper is using a practice-based framework to discuss the results to identify the opportunities and limitations of the calculator and its use as a soft policy measure. Our focus is on the perspective of the host and their experience of using the calculator. Findings from our research highlight potential pitfalls, as well as opportunities that should be further studied from a user perspective. We propose that to better understand the potential of calculators as a policy measure, future studies should take a closer look at how calculators are or can be incorporated into people's everyday practices.

Carbon Footprint in Indonesia
*DOI: 10.5614/xxxx*